\documentclass[a4paper,11pt]{article}
\pdfoutput=1
\usepackage{amsmath,amssymb}
\usepackage{IEEEtrantools}
\usepackage{setspace}
\DeclareMathOperator*{\argmax}{arg\,max}
\usepackage{graphicx}
\usepackage{cite}
\usepackage{arxiv}
\usepackage{hyperref}
\usepackage{mathtools}
\usepackage[utf8]{inputenc}
\usepackage[numbib,notlof,notlot,nottoc]{tocbibind}

\usepackage[
singlelinecheck=false % <-- important
]{caption}
\begin{document}
\setstretch{1.5}
\title{Maximum likelihood analysis of non-equilibrium solution-based single-molecule FRET data}
\author{Marijn de Boer \\ Groningen Biomolecular Sciences and Biotechnology Institute \\ University of Groningen \\ Groningen, The Netherlands \\ \href{mailto:marijndeboer4@gmail.com}{\nolinkurl{marijndeboer4@gmail.com}}}
\date{}
\maketitle

\begin{abstract}
Measuring the F\"{o}rster resonance energy transfer (FRET) efficiency of freely diffusing single molecules provides information about the sampled conformational states of the molecules. Under equilibrium conditions, the distribution of the conformational states is independent of time, whereas it can vary over time under non-equilibrium conditions. In this work,  we consider the problem of parameter inference on non-equilibrium solution-based single-molecule FRET data. With a non-equilibrium model for the conformational dynamics and a model for the conformation-dependent FRET efficiency distribution, the likelihood function could be constructed. The model parameters, such as the rate constants of the non-equilibrium conformational dynamics model and the average FRET efficiencies of the different conformational states, have been estimated from the data by maximizing the appropriate likelihood function via the Expectation-Maximization algorithm. We illustrate the likelihood method for a few simple non-equilibrium models and validated the method by simulations. The likelihood method could be applied to study protein folding, macromolecular complex formation, protein conformational dynamics and other non-equilibrium processes at the single-molecule level and in solution.
\end{abstract}


\section{Introduction}
Single-molecule methods have emerged as a powerful tool to study biological processes on the molecular level \cite{pmid28869217,pmid25264779,pmid24565277, pmid27979907, pmid29348210}. Compared to bulk methods, at which populations of molecules are observed, single-molecule methods allow one to observe individual molecules one after another. The advantage of observing individual molecules is that information about the distribution can be obtained, rather than only providing information about the population average. However, the disadvantage of single-molecule methods is that only a subsample of the population of molecules is seen, and thus, to draw any meaningful conclusion about the entire population, statistically correct inference is needed. In this work, we look at the problem of obtaining parameter estimates from single-molecule F\"{o}rster resonance energy transfer (smFRET) measurements \cite{pmid29348210}. 

In an smFRET measurement, a molecule is labeled with a donor and an acceptor dye and the energy transfer from the donor to the acceptor is measured \cite{Ha1996} (Figure 1A). The efficiency of energy transfer, termed the FRET efficiency $E$, can be determined by measuring the donor and acceptor emission intensities or by measuring the lifetime of the excited state of the donor dye \cite{lakowicz}. Importantly, $E$ depends on the distance between the donor and acceptor dye $r$ via $E =  R_0^6 / (R_0^6 + r^6 )$, where $R_0$ is a constant called the F\"{o}rster distance \cite{forster, stryer}. $E$ is most sensitive to distance changes around $R_0$. Commonly used donor and acceptor pairs have $R_0$ values between 4 and 6 nm. Thus, the sensitivity range of FRET matches with the typical size of biomolecules, making smFRET ideal to use as a 'spectroscopic ruler' \cite{stryer} to study distances between or within proteins \cite{deboerelife, Ha1999, deboerbiop, pmid29915236, pmid11087856}, nucleic acids \cite{pmid10856219, fijen} or other molecules. The distance information can be used to uncover the sampled conformational states of the molecule and the rates of interconversion between these states \cite{pmid29348210, IC}.

\begin{figure}[!h]
\centering
\includegraphics[scale = 1.0]{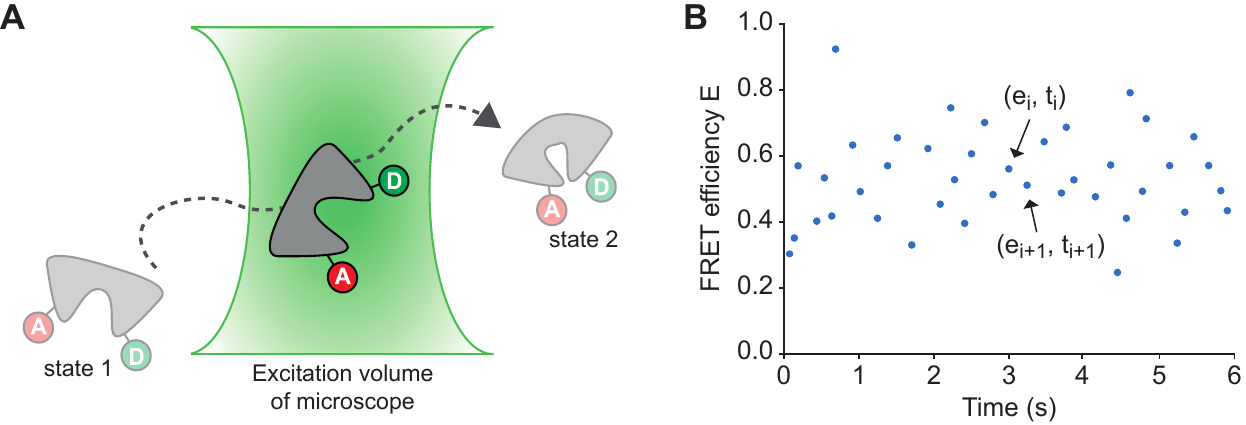}
\caption{Principle of solution-based smFRET. (A) Schematic of a protein molecule labeled with a donor and acceptor dye diffusing through the excitation volume of a focused laser beam. When the molecule is in the excitation volume, the donor is excited and the energy that is transferred to the acceptor can be measured. The molecule in this example can be in two conformations, denoted by state 1 and 2. In state 1, the donor-acceptor distance is larger than in state 2. Hence, the average FRET efficiency of state 1 is lower than of state 2. (B) Snapshot of typical solution-based smFRET data. The FRET efficiency and the time the molecule is in the excitation volume can be determined experimentally. Each blue dot in the figure corresponds to a single-molecule observation.}
\label{fig1}
\end{figure}

In a solution-based smFRET measurement (also termed diffusion-based smFRET or freely-diffusion smFRET) a very low concentration (1- 50 pM) of donor and acceptor labeled molecules is used \cite{pmid10097095} (Figure 1). The low concentration ensures that at most only one molecule is present in the excitation volume, which is created by the focused laser beam of a confocal microscope. During the diffusional transit through the excitation volume, the donor is excited, and the labeled molecule generates a short burst of photons. The detected photons can the used for the determination of $E$. In an alternating laser excitation (ALEX) \cite{alex} or pulsed interleaved excitation (PIE) \cite{pie} scheme, the presence of the acceptor dye is checked by using an additional laser.  Because of the short observation time on each individual molecule (typically around 1 ms), only a limited number of photons can be detected (typically tens or hundreds of photons per burst) leading to large uncertainties in the measured $E$ values. For correct statistical inference, the $E$ values of a large number of molecules, typically hundreds or thousands, are determined in a single measurement. When the measurement is performed under equilibrium conditions (Figure 2A), a histogram for the $E$ values can be made and fitted with a mixture of multiple Gaussian distributions (e.g. \cite{abce1}), for which each Gaussian distribution reflects a conformational state of the molecule. That is, in the mixture model, the mean of the Gaussian distribution represents the donor-acceptor distance of the conformational state and the mixture weight corresponds to its relative abundance in the population. When the measurement is performed under non-equilibrium conditions (Figure 2B), the $E$ distribution could vary over time, and fitting with a time-\textit{independent} mixture model would, in general, be inappropriate. 

Some examples of non-equilibrium processes that have been studied with solution-based smFRET are: the association and dissociation of the nucleosome core particle (NCP) \cite{NCP}, the dissociation of the ribosome recycling factor ATP-binding cassette E1 (ABCE1) from the small ribosomal subunit \cite{abce1}, promoter escape of the bacterial RNA polymerase (RNAP) \cite{pmid31356875} and the stepping action of molecular motors \cite{dna_motor, dna_motor2}. In these measurements, the system was brought out of equilibrium by diluting the sample into a new buffer environment, in which also the measurement is performed. Subsequently, fluorescence bursts were detected while the molecules diffuse through the excitation volume and they relaxes back to equilibrium.   

In this paper, we provide an analysis framework for the analysis of non-equilibrium solution-based smFRET data. Given a non-equilibrium model for the conformational dynamics and a model for the conformation-dependent FRET efficiency distribution, the likelihood function could be constructed. The likelihood function can be used for various statistical inference problems, including finding the best model parameters that describe the data. The model parameters, such as the rate constants of the non-equilibrium model and the average FRET efficiencies of the different conformational states, were found by maximizing the likelihood function. In this paper, we will provide the formulation of this likelihood function and we show that the parameter inference problem can be solved efficiently with the Expectation-Maximization (EM) algorithm \cite{EM}.

\begin{figure}[!t]
\centering
\includegraphics[width=1\textwidth]{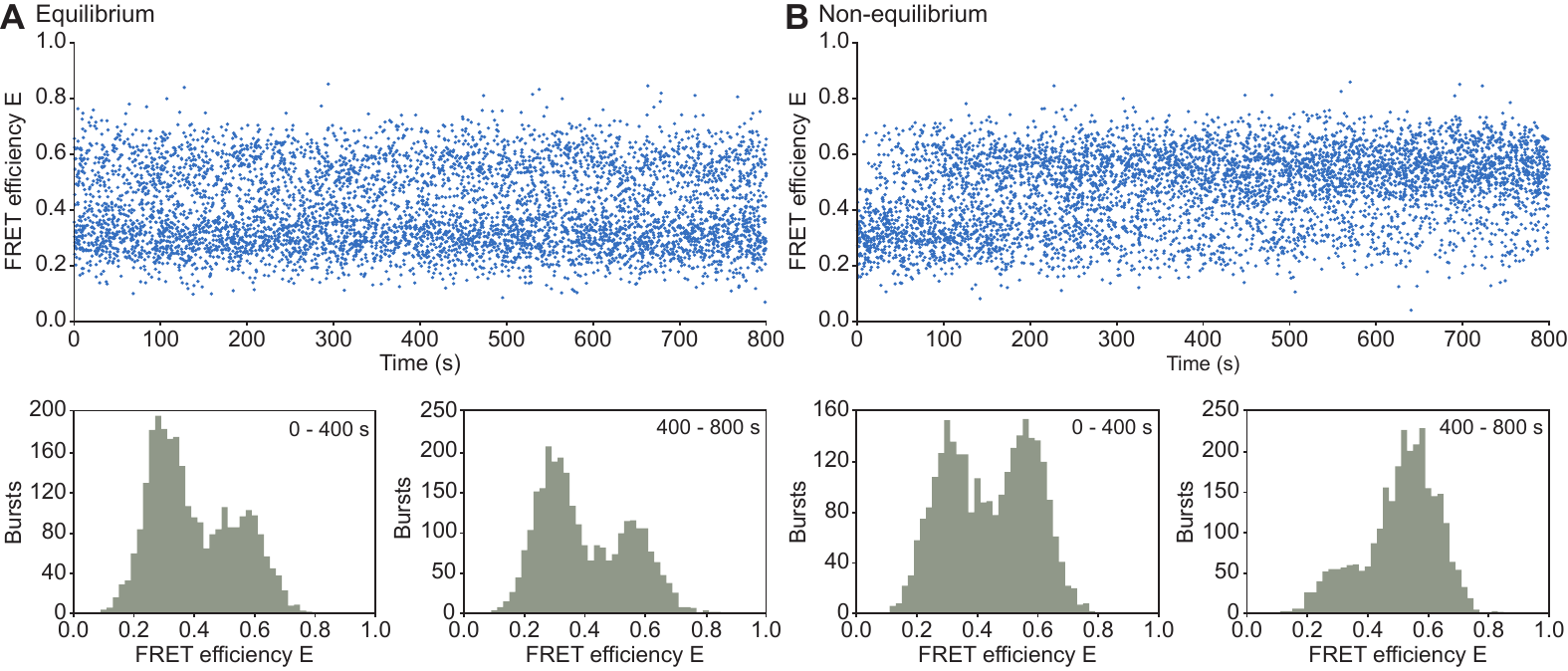}
\caption{Equilibrium and non-equilibrium solution-based smFRET. (A) Example of equilibrium solution-based smFRET data. For each molecule (blue dots) the FRET efficiency and detection time is recorded (top). The FRET efficiency histograms for bursts detected within the initial 400 s and final 400 s are depicted (bottom). (B) Same as in (A) only with non-equilibrium data. Note that in contrast to the equilibrium case, in the FRET efficiency histograms of the non-equilibrium data, substantial changes between the initial 400s and final 400 s are seen.}
\label{fig2}
\end{figure}

\section{Model definition}
In this section, we define the quantities of our model for a solution-based smFRET measurement. Suppose a measurement starts at time $0$ and stops at time $T$ ($T>0$). During this period a random number of molecules are detected while they diffuse through the excitation volume of a focused laser beam. We denote the number of detected molecules by the random variable $N(T)$. $N(T)$ is a counting process, meaning that $N(T)$ is non-negative, integer valued and non-decreasing in $T$. For notational convenience we write $N(T)$ simply as $N$ and understand its explicit dependence on $T$. We label each detected molecule by $i$ ($i \in \{1,\dots,N\}$) based on its detection time as follows. The times the $N$ molecules are detected is given by the sequence of random variables $T_1,\dots,T_N$, where $T_i$ belongs to molecule $i$ and are ordered according to $0 \leq T_1< \cdots <T_N \leq T$. We assume throughout the work that the time spend in the excitation volume (typically 1 ms) is short compared to the conformational dynamics of the molecule. Thus, if $T_i^e$ and $T_i^l$ are the time the molecule enters and leaves the excitation volume, respectively, then $T_i$ can be defined anywhere in between $T_i^e$ and $T_i^l$ (e.g. $ (T_i^l-T_i^e) /2$). We also assume that the concentration of labeled molecules is low, so that the probability of having two molecules simultaneously in the excitation volume is negligible.  

Most biomolecules, such as proteins, exist  in a countable number of conformational states that are separated by kinetic barriers \cite{boehr, IC}. Suppose we want to describe a population of identical molecules that can exist in $K$ different conformational states ($K \geq 1$), which are numbered from $1$ to $K$. The conformational state of molecule $i$ is a random variable $Z_i$ with a discrete state space ${1, \dots ,K}$. Due to measurement noise, samples from $Z_i$ are hidden to the observer and are thus the \textit{latent variables} of our model. The $N$ detected molecules give rise to the sequence $Z_1 \dots,Z_N$. We expect that the vast majority of bursts originate from different molecules, so therefore we model the $Z_i$ to be independent. In a non-equilibrium measurement, the probability that a molecule is in a particular conformational state varies over time. Therefore, the distribution of $Z_i$ is conditionally dependent on $T_i$. This property makes $Z_1, \dots ,Z_N$ a non-stationary random process. The conditional distribution of $Z_i$ given $T_i$ is denoted by $p(z | t; \theta) \coloneqq P(Z_i=z | T_i=t)$, where $z \in \{ 1,\dots,K \}$ and $\theta$ is a vector of model parameters that we want to determine from the data. For example, $\theta$ could represent a (set of) conversion rate(s) from one conformational state to another. Note that if the population of molecules is in equilibrium, then the distribution $Z_i$ is independent of time. Then $Z_1,\dots,Z_N$ is a sequence of independent and identically distributed random variables and is thus a stationary random process. 

As stated above, the conformational states of the molecules cannot be directly observed. However, different conformational states can give rise to different donor-acceptor distances, which can be determined by measuring their FRET efficiency. To avoid problems with parameter identifiability, we assume that each conformational state gives a unique donor-acceptor distance. However, the differences between the distances of the conformational states can be arbitrarily small. The FRET efficiency of molecule $i$ is denoted by the random variable $E_i$. In principle, the value of $E_i$ lies between 0 and 1. For practical reasons, this assumption is often relaxed and the range of $E_i$ is simply extended to the real line. The $N$ detected molecules give a sequence $E_1,\dots,E_N$. The $E_i$ are taken to be independent, because they are expected to originate from different molecules. In a non-equilibrium measurement, the $E_i$ are not identically distributed (see below), so $E_1,\dots,E_N$ is a non-stationary random process.

We define the distribution of $E_i$ conditional on $Z_i$ by $p(e | z; \eta_z ) \coloneqq P(E_i=e | Z_i=z)$, where $\eta_z$ is a vector of parameters that we want to infer from the data. In general, a good approximation for this conformation-dependent FRET efficiency distribution is a Gaussian distribution, which is parameterized by its mean $\mu_z$ and standard deviation $\sigma_z$. The parameter vector $\eta_z$ is in this case $\eta_z=\{ \mu_z,\sigma_z \}$. In the Gaussian approximation, $\mu_z$ represents the average FRET efficiency of the $z$th conformational state and is related to the donor-acceptor distance via $\mu_z =  R_0^6 / (R_0^6+r^6 )$. The $\sigma_z$ determines the width of the Gaussian distribution and is influenced by the measurement noise (typically $\sigma_z \approx 0.05-0.10$). In principle other distributions could also be considered, such as a beta distribution \cite{beta}, but then the computation of the maximum likelihood estimate for $\eta_z$ is more complex. 

We define the distribution of $E_i$ conditional on $T_i$ by $p(e|t; \Omega) \coloneqq P(E_i=e|T_i=t)$, where $\Omega$ is a vector of parameters we define below. By using the sum rule of probability, we obtain 
\begin{equation}
p(e|t; \Omega) = \sum_{z=1}^{K} P(E_i=e,Z_i=z|T_i=t)
\end{equation}
By using the fact that the FRET efficiency depends on the donor-acceptor distance (and $R_0$), and thus on the conformational state of the molecule, and not directly on time, we get
\begin{equation}
\begin{aligned}
p(e|t; \Omega) = & \sum_{z=1}^{K} P(E_i=e|Z_i=z,T_i=t) P(Z_i=z|T_i=t) \\
= & \sum_{z=1}^{K} P(E_i=e|Z_i=z) P(Z_i=z|T_i=t) 
\end{aligned}
\end{equation}
so
\begin{equation}
p(e|t; \Omega) = \sum_{z=1}^{K} p(e | z ; \eta_z) p(z | t ; \theta)  \label{mixture}
\end{equation}
From Eq. \ref{mixture} it follows that the parameter vector $\Omega$ is $\{\theta,\eta_1,…,\eta_K\}$. $p(e|t; \Omega)$ gives the probability that a molecule detected at time $t$ has a FRET efficiency of $e$, which, according to Eq. \ref{mixture}, is equal to the sum of probabilities that the FRET efficiency is generated by the different conformational states (at time $t$). An example of $p(e|t; \Omega)$, $p(e | z ; \eta_z)$ and $p(z | t ; \theta)$ is given in Figure 3. We will further elaborate on how to deduce the model parameters from the measurement data and after that, a few examples will be given for illustration purposes.

\begin{figure}[!h]
\centering
\includegraphics[scale = 0.95]{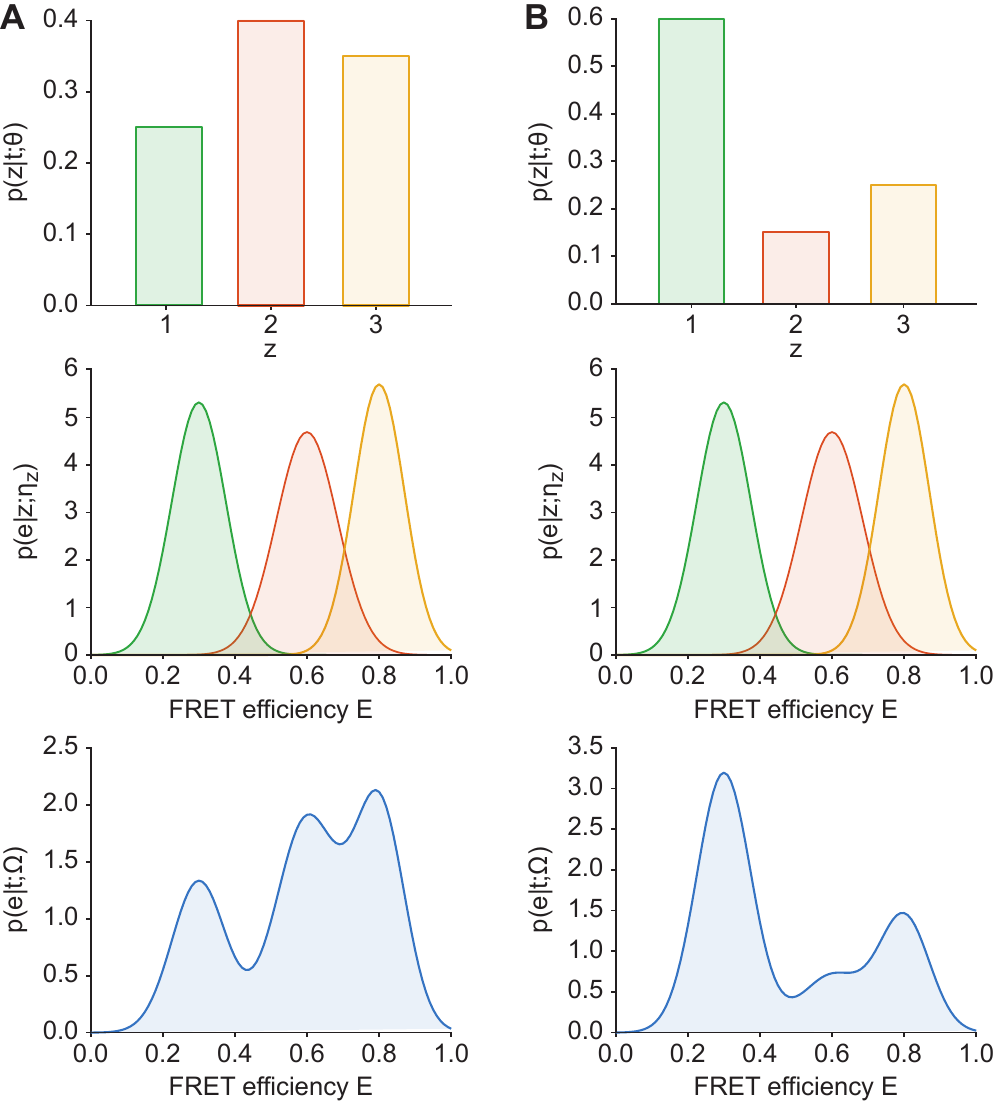}
\caption{Example of distributions $p(z | t ; \theta)$ (top), $p(e | z ; \eta_z)$ (middle) and $p(e|t; \Omega)$ (bottom) with three conformational states ($K=3$) at one (A) and at another timepoint (B). $p(e|t; \Omega)$ is calculated with Eq. \ref{mixture} and the $p(e | z ; \eta_z)$ are Gaussian.}
\label{figX}
\end{figure}

\section{Likelihood estimation}
We use the method of maximum likelihood estimation to infer the model parameters from the data. Maximum likelihood estimation is a statistical inference method that has often been applied in the analysis of single-molecule data \cite{pmid25088495, pmid19588948, pmid16679362, pmid19626138, pmid6311301, pmid18266353}. From the model, the likelihood function can be constructed and maximized with respect to the unknown model parameters. This yields the maximum likelihood estimates for the parameters, which have the desired property that they converge (almost surely) to the true parameters if the amount of data becomes large and if the statistical model is correct. 

The dataset of a solution-based smFRET measurement can be represented by $D=\{(e_1,t_1 ),…,(e_n,t_n)\}$, where $e_i$, $t_i$ and $n$ are a sample of $E_i$, $T_i$ and $N$, respectively (Figure 1B). The parameters we want to estimate from the dataset are represented by $\Omega$.

By using the fact that the observations on each molecule are independent, the likelihood function can be expressed as the following product
\begin{equation}
\begin{aligned}
L(\Omega ; D)' = & P(E_1=e_1, \dots,E_n=e_n,T_1=t_1, \dots,T_n=t_n;\Omega) \\
= & P(T_1=t_1, \dots,T_n=t_n) \prod_{i=1}^n p(e_i | t_i ; \Omega )
\end{aligned}
\end{equation}
where $P(T_1=t_1, \dots,T_n=t_n)$ is the probability of detecting the $n$ molecules at time $t_1,\dots,t_n$. Because this probability does not depend on $\Omega$, we can work with the conditional likelihood
\begin{equation}
L(\Omega ; D) = \prod_{i=1}^{n} p(e_i | t_i ; \Omega ) = \prod_{i=1}^{n} \sum_{z=1}^{K} p(e_i | z ; \eta_z) p(z | t_i ; \theta)
\end{equation}
where for the last step we used Eq. \ref{mixture}. For computational convenience, we take the natural logarithm of $L(\Omega;D)$ 
\begin{equation}
l(\Omega ; D) = \log L(\Omega ; D)  = \sum_{i=1}^{n} \log \sum_{z=1}^{K} p(e_i | z ; \eta_z) p(z | t_i ; \theta) \label{logLL}
\end{equation}
The maximum likelihood estimates for $\theta$ and $\eta_1,…,\eta_K$ are found from maximizing $l(\Omega ; D)$
\begin{equation}
\hat{\Omega}  =  \argmax_{\Omega} l(\Omega ; D)  \label{argmaxlikelihood}
\end{equation}
where $\hat{\Omega} = \{\hat{\theta}, \hat{\eta}_z,\dots,\hat{\eta}_K \}$. Unfortunately, the maximization of $l(\Omega;D)$ is, in general, quite difficult because of the summation inside the logarithm. We will show that the maximization can often be simplified when the EM algorithm is used \cite{EM}. The EM algorithm is an iterative method that can simplify the optimization problem when the model involves latent variables (e.g. as in hidden Markov models \cite{hmm}).

The EM algorithm starts with an initial estimate for the parameters $\Omega^{(0)}$ that are iteratively updated via two subsequent steps, termed the Expectation (E) and Maximization (M) step. After $k$ iterations, we use our current estimate of $\Omega$, i.e. $\Omega^{(k)}=\{\theta^{(k)},\eta_1^{(k)},…,\eta_K^{(k)} \}$, in the E step to calculate
\begin{equation}
\begin{aligned} \label{Q}
Q(\Omega | \Omega^{(k)}) = &\mathbb{E}_{\boldsymbol{Z} | \boldsymbol{E}, \boldsymbol{T}, \Omega^{(k)}} [l(\Omega;\boldsymbol{Z},\boldsymbol{E}|\boldsymbol{T})] \\
= & \sum_{i=1}^{n} \sum_{z=1}^{K} p(z | e_i , t_i; \Omega^{(k)}) \log{p(e_i,z | t_i; \Omega)}  \\
= & \sum_{i=1}^{n} \sum_{z=1}^{K} p(z | e_i , t_i; \Omega^{(k)}) \log{ \left( p(e_i | z; \eta_z) p(z|t_i ; \theta) \right) }
\end{aligned}
\end{equation}
where $l(\Omega;\boldsymbol{Z},\boldsymbol{E}|\boldsymbol{T})$ is the log-likelihood of $\boldsymbol{Z}=\{Z_1, \dots, Z_n\}$ and $\boldsymbol{E}=\{E_1, \dots, E_n\}$ conditional on $\boldsymbol{T}=\{T_1, \dots, T_n\}$, and $\mathbb{E}_{\boldsymbol{Z} | \boldsymbol{E},\boldsymbol{T}, \Omega^{(k)}}[\cdot]$ denotes the expectation with respect to the conditional distribution of $\boldsymbol{Z}$ given $\boldsymbol{E}$, $\boldsymbol{T}$ and the current estimates of the parameters $\Omega^{(k)}$. In Eq. \ref{Q}, $p(e_i | z; \eta_z )$ and $p(z| t_i ; \theta)$ are the models for the conformation-dependent FRET efficiency and the non-equilibrium conformational dynamics, respectively (Section 2), and $p(z | e_i,t_i; \Omega^{(k)})$ is calculated with Bayes rule 
\begin{equation}
p(z | e_i,t_i; \Omega^{(k)}) = \frac{p(e_i | z; \eta_z^{(k)}) p(z|t_i ; \theta^{(k)})}{\sum_{z=1}^{K} p(e_i | z; \eta_z^{(k)}) p(z|t_i ; \theta^{(k)})} \label{bayes}
\end{equation}
In the M step, the parameters are updated by maximizing $Q(\Omega | \Omega^{(k)})$ with respect to $\Omega$
\begin{equation}
\Omega^{(k+1)}  =  \argmax_{\Omega} Q(\Omega | \Omega^{(k)}) \label{argmaxQ}
\end{equation}
The E and M step are repeated until convergence is reached. One reasonable convergence test would be to stop iterating when $Q(\Omega^{(k+1)} | \Omega^{(k)}) - Q(\Omega^{(k)} | \Omega^{(k)}) < \epsilon$, where $\epsilon$ is a threshold value. 

The optimization problem of Eq. \ref{argmaxQ} can be expressed as separate optimization problems, because $Q(\Omega|\Omega^{(k)})$ can be expressed as the following sum
\begin{equation}
Q(\Omega | \Omega^{(k)}) = h(\theta | \Omega^{(k)}) + \sum_{z=1}^{K} g(\eta_z | \Omega^{(k)}) \label{expansion}
\end{equation}
where
\begin{equation}
h(\theta | \Omega^{(k)}) = \sum_{i=1}^{n} \sum_{z=1}^{K} p(z | e_i , t_i; \Omega^{(k)}) \log{p(z|t_i ; \theta)} \label{h}
\end{equation}
and
\begin{equation}
g(\eta_z | \Omega^{(k)}) = \sum_{i=1}^{n} p(z | e_i , t_i; \Omega^{(k)}) \log{p(e_i | z; \eta_z)}  \label{g}
\end{equation}
Thus, the parameters that maximize $Q(\Omega|\Omega^{(k)})$ are found from maximizing each of the individual terms in Eq. \ref{expansion}
\begin{equation}
\theta^{(k+1)}  =  \argmax_{\theta} h(\theta| \Omega^{(k)}) 
\end{equation}
and
\begin{equation}
\eta_z^{(k+1)}  =  \argmax_{\eta_z} g(\eta_z| \Omega^{(k)}) 
\end{equation}
for $z \in \{1,\dots, K \}$.

Up to this point, we did not make any explicit assumption about the distributions $p(e_i |z; \eta_z )$ and $p(z|t_i; \theta)$, and is thus applicable to any model for the conformation-dependent FRET efficiency and the non-equilibrium conformational dynamics. It shows how the maximum likelihood estimates can be found with the EM algorithm, and that the optimization problem can be expressed as $K+1$ smaller optimization tasks involving the separate optimization for $\theta$ and the $K$ $\eta_z$ parameter vectors. This result is important from a computational perspective. Without the EM algorithm, the maximization of the likelihood function of Eq. \ref{argmaxlikelihood} involves $K+1$  parameter vectors. However, by using the EM algorithm, the maximum is found by solving $K+1$ separate optimization problems, which, in general, drastically simplifies the problem.

\section{Gaussian FRET efficiency distribution}
Before we look at some illustrative examples of the above analysis framework, we first consider how the M step can be implemented in case the conformation-dependent FRET efficiency distribution is Gaussian. In this case, $p(e_i|z;\mu_z, \sigma_z) = \mathcal{N}(e_i ; \mu_z, \sigma_z)$, where
\begin{equation}
\mathcal{N}(e_i ; \mu_z, \sigma_z) = \frac{1}{\sqrt{2 \pi \sigma_z^2}} e^{-\frac{(e_i - \mu_z)^2}{2 \sigma_z^2}} \label{normal_distribution}
\end{equation}
The Gaussian distribution is parameterized by its mean $\mu_z$ and standard deviation $\sigma_z$, so $\eta_z=\{\mu_z,\sigma_z\}$. As long as the FRET efficiency does not lie close to 0 or 1, the Gaussian approximation is often a very good approximation of the FRET efficiency distribution (of a single distance). 

In the M step, we need to maximize $g(\eta_z| \Omega^{(k)}) = g(\mu_z, \sigma_z| \Omega^{(k)})$ with respect to $\mu_z$ and $\sigma_z$. From Eq. \ref{g} we have
\begin{equation}
\begin{aligned}
g(\mu_z, \sigma_z | \Omega^{(k)})  = &  \sum_{i=1}^{n} p(z | e_i , t_i; \Omega^{(k)}) \log{\mathcal{N}(e_i ; \mu_z, \sigma_z)}   \\  \label{gauss}
= & - \sum_{i=1}^{n} p(z | e_i , t_i; \Omega^{(k)}) \left( \frac{(e_i - \mu_z)^2}{2 \sigma_z^2} + \log{\sqrt{2 \pi \sigma_z^2}} \right)
\end{aligned}
\end{equation}
In Eq. \ref{gauss}, $p(z|e_i,t_i;\Omega^{(k)})$  is determined during the E step via Eq. \ref{bayes} and will be considered in more detail in the next section. Maximizing Eq. \ref{gauss} with respect to $\mu_z$ and $\sigma_z$ gives
\begin{equation}
\mu_z^{(k+1)} = \frac{\sum_{i=1}^n p(z|e_i,t_i;\Omega^{(k)}) e_i}{\sum_{i=1}^n p(z|e_i,t_i;\Omega^{(k)})} \label{Mstep_mu}
\end{equation}
and 
\begin{equation}
\sigma_z^{(k+1)} = \sqrt{\frac{\sum_{i=1}^n p(z|e_i,t_i;\Omega^{(k)}) \left( e_i - \mu_z^{(k+1)}\right)^2}{\sum_{i=1}^n p(z|e_i,t_i;\Omega^{(k)})}} \label{Mstep_sigma}
\end{equation}
In conclusion, the parameters $\mu_z$ and $\sigma_z$ are updated during the M step by determining the weighted sample average and the weighted sample standard deviation of the measured FRET efficiencies.

\section{Illustrative examples}
In this section we consider simple examples of models for non-equilibrium conformational dynamics, with a focus on the calculation of the E and M step. Throughout this section we assume that the conformation-dependent FRET efficiencies are Gaussian, and its implementation for the M step was covered in the previous section.

\subsection{Example 1: Exponential two-state model}
Suppose a molecule can be in two conformations, denoted by state 1 and state 2. At the beginning of the measurement, all the molecules start  in state 1 and go randomly to state 2 with a rate constant $\lambda$. If the rate constant is independent of time, then the lifetime of state 1 is exponentially distributed. Exponential lifetimes are commonly observed in smFRET data \cite{deboerelife, pmid24758940, pmid18026086, pmid22669904, pmid30272207, pmid23502425}. We assume that molecules in state 2 cannot convert back to state 1. This process could, for example, correspond to the folding of a protein molecule. In this process, all molecules are initially in the unfolded state (state 1) and fold with a rate constant $\lambda$ to the folded state (state 2). In this particular example, the process is irreversible, so the folded molecule cannot convert back to the unfolded state. Under these conditions, the model for the conformational dynamics is 
\begin{equation}
p(z | t; \lambda) = \left\{
\begin{array}{cc}
e^{-\lambda t} &  z = 1 \\
1 - e^{-\lambda t} & z = 2  \label{model1}
\end{array} \right.
\end{equation}
with $\lambda>0$. In this example the $K+1$ vectors (where $K = 2$) are $\theta=\{\lambda\}$, $\eta_1=\{\mu_1,\sigma_1 \}$ and $\eta_2=\{\mu_2,\sigma_2 \}$. And thus the set of model parameters is $\Omega=\{\lambda,\mu_1,\sigma_1,\mu_2,\sigma_2 \}$. In Figure 4A, $p(z | t; \lambda)$ is plotted for $\lambda =0.01$ s$^{-1}$, and in Figure 4B, $p(e | t; \Omega)$ is plotted by using Eq. \ref{mixture}, where $p(z | t; \eta_z)$ is a Gaussian distribution with $\mu_1=0.5$, $\sigma_1=0.07$, $\mu_2=0.7$ and $\sigma_2=0.06$.

We can use the EM algorithm to find the maximum likelihood estimates for $\Omega$. For the E step calculation in Eq. \ref{bayes}, we use Eq. \ref{model1} constructed with our current parameter estimates $\lambda^{(k)}$, $\mu_1^{(k)}$, $\sigma_1^{(k)}$, $\mu_2^{(k)}$ and $\sigma_2^{(k)}$ to calculate 
\begin{equation}
p(z | e_i, t_i; \Omega^{(k)}) = \left\{
\begin{array}{cc}
\frac{ e^{-\lambda^{(k)} t_i} \mathcal{N}(e_i ; \mu_1^{(k)}, \sigma_1^{(k)}) }{ e^{-\lambda^{(k)} t_i } \mathcal{N} (e_i ; \mu_1^{(k)}, \sigma_1^{(k)}) + \left( 1 - e^{-\lambda^{(k)} t_i} \right) \mathcal{N}(e_i ; \mu_2^{(k)}, \sigma_2^{(k)}) } &  z = 1 \\
\frac{ \left( 1 - e^{-\lambda^{(k)} t_i} \right) \mathcal{N}(e_i ; \mu_2^{(k)}, \sigma_2^{(k)}) }{ e^{-\lambda^{(k)} t_i } \mathcal{N}(e_i ; \mu_1^{(k)}, \sigma_1^{(k)}) + \left( 1 - e^{-\lambda^{(k)} t_i} \right) \mathcal{N}(e_i ; \mu_2^{(k)}, \sigma_2^{(k)}) } & z = 2 \label{Esetp_model1}
\end{array} \right.
\end{equation}
where $\mathcal{N}(e_i ; \mu_1^{(k)}, \sigma_1^{(k)})$ and $\mathcal{N}(e_i ; \mu_2^{(k)}, \sigma_2^{(k)})$ are defined by Eq. \ref{normal_distribution}. With $p(z|e_i,t_i; \Omega^{(k)} )$ calculated, we proceed to the M step and update the parameters. The M step for $\mu_1$, $\sigma_1$, $\mu_2$ and $\sigma_2$ are given by Eq. \ref{Mstep_mu} and \ref{Mstep_sigma} and using $p(z|e_i,t_i;\Omega^{(k)} )$ as given by Eq. \ref{Esetp_model1}. To update $\lambda$, we use Eq. \ref{h}, \ref{model1} and \ref{Esetp_model1} to obtain
\begin{equation}
\begin{aligned}
h(\lambda | \Omega^{(k)})  = &   \sum_{i=1}^{n} \sum_{z=1}^{K} p(z | e_i , t_i; \Omega^{(k)}) \log{p(z|t_i ; \lambda)}  \\  \label{estep_model1b}
= & \sum_{i=1}^{n}  - p(z = 1| e_i , t_i ; \Omega^{(k)}) \lambda t_i + p(z = 2| e_i , t_i ; \Omega^{(k)}) \log \left( 1  - e^{-\lambda t_i} \right)
\end{aligned}
\end{equation}
To give the next iteration value $\lambda^{(k+1)}$, Eq. \ref{estep_model1b} needs to be maximized with respect to $\lambda$. No closed form expression for $\lambda^{(k+1)}$ exists. Fortunately, $h(\lambda | \Omega^{(k)})$ is a concave function of $\lambda$, because
\begin{equation}
\frac{\partial^2 h(\lambda | \Omega^{(k)})}{\partial^2 \lambda} = - \sum_{i=1}^{n} \frac{p(z = 2| e_i , t_i ; \Omega^{(k)}) t_i^2 e^{\lambda t_i}}{\left( e^{\lambda t_i} -1 \right)^2} < 0
\end{equation}
for every $\lambda>0$, so the maximum of $h(\lambda | \Omega^{(k)})$ can be found relatively straightforwardly with any numerical optimization algorithm. 

Note that in this particular example, the multi-parameter optimization problem of Eq. \ref{argmaxlikelihood} involves 5 parameters and is reduced by the EM algorithm to 3 separate and simpler optimization problems. 

\begin{figure}[!h]
\centering
\includegraphics[scale = 1]{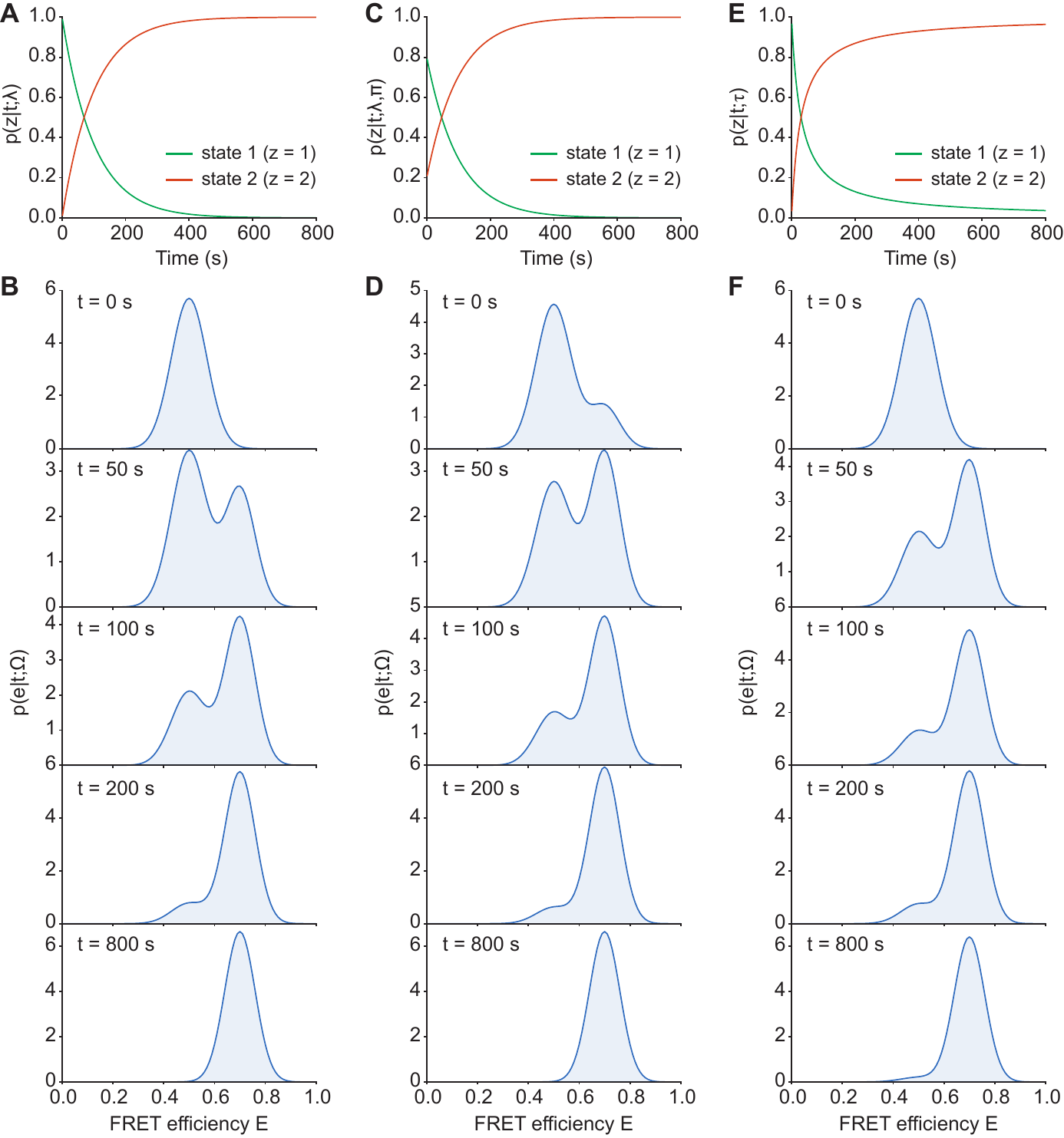}
\caption{Example of non-equilibrium conformational dynamics models. Eq. \ref{model1} with $\lambda=0.01$ s$^{-1}$ (A), Eq. \ref{model2} with $\lambda=0.01$ s$^{-1}$ and $\pi=0.8$ (C) and Eq. \ref{model3} with $\tau=100$ s (E) are plotted. FRET efficiency distribution at different timepoints, when the conformational dynamics is modeled according to panel A (B), C (D) and E (F). The conformation-dependent FRET efficiency distribution is Gaussian with $\mu_1=0.5$, $\sigma_1=0.07$ for state 1 and $\mu_2=0.7$ and $\sigma_2=0.06$ for state 2.}
\label{fig3}
\end{figure}

\subsection{Example 2: Exponential two-state model with offset}
We modify example 1, by assuming that at the beginning of the measurement some molecules within the population are already in state 2. Let $\pi$ denote the fraction of molecules being in state 1 at the beginning of the measurement and $1-\pi$ the fraction of molecules in state 2. Then the model for the conformational dynamics is 
\begin{equation}
p(z | t; \lambda,\pi) = \left\{
\begin{array}{cc}
\pi e^{-\lambda t} &  z = 1 \\
1 - \pi e^{-\lambda t} & z = 2  \label{model2}
\end{array} \right.
\end{equation}
with $\lambda > 0$ and $0<\pi<1$.  In this example, $\theta=\{\lambda,\pi \}$, $\eta _z=\{\mu_z,\sigma \}$ and $\Omega=\{\lambda,\pi,\mu_1,\sigma_1,\mu_2,\sigma_2 \}$.  In Figure 4C, $p(z | t; \lambda)$ is plotted for $\lambda =0.01$ s$^{-1}$ and $\pi=0.8$.  In Figure 4D, $p(e | t; \Omega)$ is plotted by using Eq. \ref{mixture}, where $p(z | t; \eta_z)$ is a Gaussian distribution with $\mu_1=0.5$, $\sigma_1=0.07$, $\mu_2=0.7$ and $\sigma_2=0.06$.

We use the EM algorithm to find the maximum likelihood estimate for $\Omega$. During the E step, we use Eq. \ref{bayes} and \ref{model2} together with our current parameter estimates $\lambda^{(k)}$, $\pi^{(k)}$, $\mu_1^{(k)}$, $\sigma_1^{(k)}$, $\mu_2^{(k)}$ and $\sigma_2^{(k)}$, to calculate
\begin{equation}
p(z | e_i, t_i; \Omega^{(k)}) = \left\{
\begin{array}{cc}
\frac{ \pi^{(k)} e^{-\lambda^{(k)} t_i} \mathcal{N}(e_i ; \mu_1^{(k)}, \sigma_1^{(k)}) }{ \pi^{(k)} e^{-\lambda^{(k)} t_i } \mathcal{N}(e_i ; \mu_1^{(k)}, \sigma_1^{(k)}) + \left( 1 - \pi^{(k)} e^{-\lambda^{(k)} t_i} \right) \mathcal{N}(e_i ; \mu_2^{(k)}, \sigma_2^{(k)}) } &  z = 1 \\
\frac{ \left( 1 - \pi^{(k)} e^{-\lambda^{(k)} t_i} \right) \mathcal{N}(e_i ; \mu_2^{(k)}, \sigma_2^{(k)}) }{ \pi^{(k)} e^{-\lambda^{(k)} t_i } \mathcal{N}(e_i ; \mu_1^{(k)}, \sigma_1^{(k)}) + \left( 1 - \pi^{(k)}e^{-\lambda^{(k)} t_i} \right) \mathcal{N}(e_i ; \mu_2^{(k)}, \sigma_2^{(k)}) } & z = 2 \label{Esetp_model2}
\end{array} \right.
\end{equation}
The M step for $\mu_1$ ,$\sigma_1$, $\mu_2$ and $\sigma_2$ are given by Eq. \ref{Mstep_mu} and \ref{Mstep_sigma} and using $p(z|e_i,t_i;\Omega^{(k)} )$ as given by Eq. \ref{Esetp_model2}. To update $\lambda$ and $\pi$, we use Eq. \ref{h}, \ref{model2} and \ref{Esetp_model2} to calculate
\begin{equation}
\begin{aligned}
h(\lambda, \pi | \Omega^{(k)})  = &   \sum_{i=1}^{n} \sum_{z=1}^{K} p(z | e_i , t_i; \Omega^{(k)}) \log{p(z|t_i ; \lambda,\pi)}  \\ 
= & \sum_{i=1}^{n} p(z = 1| e_i , t_i ; \Omega^{(k)}) \left( \log \pi - \lambda t_i  \right)  + p(z = 2| e_i , t_i ; \Omega^{(k)}) \log \left( 1  - \pi e^{-\lambda t_i} \right)
\end{aligned}
\end{equation}
which can be maximized numerically to give $\lambda^{(k+1)}$ and $\pi^{(k+1)}$. Note that the complex multi-parameter optimization problem of Eq. \ref{argmaxlikelihood} is reduced by the EM algorithm to 3 separate and simpler optimization problems.

\subsection{Example 3: Non-exponential two-state model}
Similar as in the previous two examples, we consider here that a molecule can be in two conformational states, denoted by state 1 and state 2. At the beginning of the measurement all molecules are in state 1, and transit randomly to state 2. The process is irreversible, so once a molecule is in state 2 it cannot convert back to state 1. However, in contrast to the previously discussed examples, instead of the exponential lifetime of state 1, we consider here the following simple model for non-equilibrium conformational dynamics
\begin{equation}
p(z | t; \tau) = \left\{
\begin{array}{cc}
\frac{\tau}{\tau +t} &  z = 1 \\
\frac{t}{\tau +t} & z = 2  \label{model3}
\end{array} \right.
\end{equation}
with $\tau>0$. In this model, $\theta=\{\tau\}$, $\eta_1=\{\mu_1,\sigma_1 \}$, $\eta_2=\{\mu_2,\sigma_2 \}$ so $\Omega=\{\tau,\mu_1,\sigma_1,\mu_2,\sigma_2 \}$. The model can be interpreted as the irreversible transition from state 1 to state 2, occurring with a (hazard) rate that depends on time as $-\frac{\partial}{\partial t} \log \left( p(z = 1 | t; \tau) \right) = 1/(\tau + t)$. Thus, the transition becomes slower as time increases, whereas in example 1 and 2 it was independent of time. In Figure 4E, $p(z | t; \tau)$ is plotted for $\tau =100$ s, and in Figure 4F, $p(e | t; \Omega)$ is plotted by using Eq. \ref{mixture}, with $p(z | t; \eta_z)$ being Gaussian with $\mu_1=0.5$, $\sigma_1=0.07$, $\mu_2=0.7$ and $\sigma_2=0.06$.

Irrespective of the precise molecular interpretation, we can use the EM algorithm to find the maximum likelihood estimate for $\Omega$. In the E step, we use our current parameter estimates $\tau^{(k)}$, $\mu_1^{(k)}$, $\sigma_1^{(k)}$, $\mu_2^{(k)}$ and $\sigma_2^{(k)}$ together with Eq. \ref{bayes} and \ref{model3} to calculate 
\begin{equation}
p(z | e_i, t_i; \Omega^{(k)}) = \left\{
\begin{array}{cc}
\frac{\tau^{(k)} \mathcal{N}(e_i ; \mu_1^{(k)}, \sigma_1^{(k)}) }{ \tau^{(k)} \mathcal{N}(e_i ; \mu_1^{(k)}, \sigma_1^{(k)}) + t_i \mathcal{N}(e_i ; \mu_2^{(k)}, \sigma_2^{(k)}) } &  z = 1 \\
\frac{t_i \mathcal{N}(e_i ; \mu_2^{(k)}, \sigma_2^{(k)}) }{ \tau^{(k)} \mathcal{N}(e_i ; \mu_1^{(k)}, \sigma_1^{(k)}) + t_i \mathcal{N}(e_i ; \mu_2^{(k)}, \sigma_2^{(k)}) } & z = 2 \label{Esetp_model3}
\end{array} \right.
\end{equation}
The M step for $\mu_1$, $\sigma_1$, $\mu_2$ and $\sigma_2$ are given by Eq. \ref{Mstep_mu} and \ref{Mstep_sigma} and using $p(z | e_i, t_i; \Omega^{(k)})$ as given by Eq. \ref{Esetp_model3}. To update $\tau$, we combine Eq. \ref{h}, \ref{model3} and \ref{Esetp_model3} to give
\begin{equation}
\begin{aligned}
h(\tau | \Omega^{(k)})  = &   \sum_{i=1}^{n} \sum_{z=1}^{K} p(z | e_i , t_i; \Omega^{(k)}) \log{p(z|t_i ; \tau)}  \\ 
= & \sum_{i=1}^{n} p(z = 1| e_i , t_i ; \Omega^{(k)}) \log{ \left( \frac{\tau}{\tau + t_i}  \right)}  + p(z = 2| e_i , t_i ; \Omega^{(k)}) \log{ \left( \frac{t_i}{\tau + t_i}  \right)} \\
= & \sum_{i=1}^{n} p(z = 1| e_i , t_i ; \Omega^{(k)}) \log \tau - \log(\tau + t_i) + p(z = 2| e_i , t_i ; \Omega^{(k)}) \log t_i
\end{aligned}
\end{equation}
Maximizing $h(\tau | \Omega^{(k)}) $ with respect to $\tau$ gives 
\begin{equation}
\tau^{(k+1)}  =  \argmax_{\tau} \left( \sum_{i=1}^{n} p(z = 1| e_i , t_i ; \Omega^{(k)}) \log \tau - \log(\tau + t_i) \right)
\end{equation}
which can be solved relatively straightforwardly with any numerical method.

\subsection{Example 4: K-state model at equilibrium}
If a population of molecules are in equilibrium, then the stochastic process is stationary. Then, the probabilities to be in one of the conformational states do not depend on time. Let in this case $p(z | t_i ; \theta )=p(z;\theta )=w_z$. 
Suppose a population of molecules are in equilibrium and can acquire $K$ conformational states. In this example $\theta=\{w_1, \dots ,w_K \}$, $\eta_z=\{\mu_z, \sigma_z\}$ and $\Omega=\{w_1, \dots,w_K,\mu_1,\dots,\mu_K,\sigma_1,\dots,\sigma_K \}$. We use the EM algorithm to find the maximum likelihood estimate for $\Omega$. During the E step, we use Eq. \ref{bayes} together with our current parameter estimates $\Omega^{(k)}$, to calculate $p(z | e_i, t_i ; \Omega^{(k)}) = p(z | e_i ; \Omega^{(k)})$, where
\begin{equation}
p(z | e_i ; \Omega^{(k)}) = \frac{w_z^{(k)} \mathcal{N}(e_i; \mu_z^{(k)}, \sigma_z^{(k)})}{\sum_{z=1}^K w_z^{(k)} \mathcal{N}(e_i; \mu_z^{(k)}, \sigma_z^{(k)}) } \label{Estep_model4}
\end{equation}
The M step for $\mu_z$ and $\sigma_z$ are given by Eq. \ref{Mstep_mu} and \ref{Mstep_sigma} and using $p(z | e_i ; \Omega^{(k)})$ as given by Eq. \ref{Estep_model4}. To update $w_z$, we combine Eq. \ref{h} and \ref{Estep_model4}
\begin{equation}
\begin{aligned}
h(w_1, \dots,w_K| \Omega^{(k)})  = & \sum_{i=1}^{n} \sum_{z=1}^{K} p(z | e_i ; \Omega^{(k)}) \log{w_z} \\  \label{h_model3}
= & \sum_{z=1}^{K} \left[ \sum_{i=1}^{n} p(z | e_i ; \Omega^{(k)}) \right] \log w_z
\end{aligned}
\end{equation}
This needs to be maximized with respect to $w_1, \dots ,w_K$ subject to the constraint $ \sum_{z=1}^{K} w_z = 1$. This optimization problem is equivalent to finding the maximum likelihood estimates of a multinomial distribution, so  
\begin{equation}
w_z^{(k+1)} = \frac{1}{n} \sum_{i=1}^n p(z | e_i ; \Omega^{(k)})
\end{equation}
It may not come as a surprise that, when the process is in equilibrium and the conformation-dependent FRET efficiency distributions are Gaussian, we retrieve the E and M step of a Gaussian Mixture Model (GMM). 

\section{Model selection}
An important part of statistical inference through maximum likelihood is the selection of the appropriate model for the data. Here, we propose a simple scheme to visualize how well the model with the estimated parameters fit the data. The idea is to compare a histogram constructed over a certain time interval with the expected distribution of the model over that interval. Other measures could also be used for model selection, such as the Bayesian information criterion (BIC) or the Akaike information criterion (AIC), but those will not be considered here.

To make our idea concrete, we first partition the time period $[0,T]$ into $m$ non-overlapping time intervals, which we denote by $\mathcal{T}_j$, $j \in \{1,\dots, m\}$.  For $j =1$, the time interval $\mathcal{T}_1$ is defined as $\mathcal{T}_1=[0,\tau_1^1]$ with $\tau_1^1>0$. For $j \in \{2,\dots, m-1\}$, the intervals are defined as $\mathcal{T}_j=(\tau_j^0,\tau_j^1]$ where $\tau_{j-1}^1=\tau_j^0<\tau_j^1=\tau_{j+1}^0$. For $j=m$, $\mathcal{T}_m=(\tau_m^0,T]$ with $\tau_m^0<T$. Next, we construct a histogram for the FRET efficiency of all molecules that have been detected within the time interval $T_j$. In Figure 2 an example is shown for $m = 2$ and the corresponding histograms over $\mathcal{T}_1=[0$ s$,400$ s$]$ and $\mathcal{T}_2=(400$ s$,800$ s$]$. We can compare this histogram with the FRET efficiency distribution over the time interval $\mathcal{T}_j$, which we denote by $p_{\mathcal{T}_j}(E)$  and is equal to
\begin{equation}
p_{\mathcal{T}_j}(E) = \frac{1}{N_j} \sum_{t_i \in \mathcal{T}_j} p(E | t_i ; \hat{\Omega} ) \label{fit1}
\end{equation}
where $E$ is the FRET efficiency and $p(E | t_i ; \hat{\Omega} )$ is given by Eq. \ref{mixture}. The normalization constant $N_j$ is equal to the number of detected molecules in $\mathcal{T}_j$ and have the property that $\sum_{j=1}^m N_j = n$. The time intervals should be chosen such that $N_j \geq 1$ for every interval. By using Eq. \ref{mixture}, Eq. \ref{fit1} can be expressed as 
\begin{equation}
p_{\mathcal{T}_j}(E) = \frac{1}{N_j} \sum_{t_i \in \mathcal{T}_j} \sum_{z=1}^{K} p(E | z ; \hat{\eta}_z) p(z | t_i ; \hat{\theta}) \label{fit2}
\end{equation}
were $\hat{\eta}_z$ and $\hat{\theta}$ are the maximum likelihood estimates of $\eta_z$ and $\theta$, respectively. Eq. \ref{fit2} depends on the selected model for the conformation-dependent FRET efficiency distribution and the selected model for the conformational dynamics, which are both parameterized by the maximum likelihood estimates as obtained from the complete dataset. The quality of the fit can now be visualized by examining the discrepancy between the histograms and the model prediction of Eq. \ref{fit2} (see next section for an example). In principle, the discrepancy could be made precise by any measure of goodness of fit, such as the chi-squared test. 

\section{Numerical calculation}
In this section, we test the maximum likelihood method on simulated data. To do this, we simulated data with typical experimental parameters and compared these parameters to the maximum likelihood estimates that are obtained from the simulated data. 

In all our simulations, we simulated the data of 3000 bursts. The timepoints the bursts are detected is modeled to be Poisson process with a rate 10 s$^{-1}$. In this case, the average time between two subsequent bursts is 100 ms and the average measurement time is 5 min ($3000 \cdot 100$ ms). The typical residence time of a molecule in the excitation volume is 1 ms, so the probability to have two molecules simultaneously in the excitation volume is $10^{-2} \cdot 10^{-2} = 10^{-4}$ and can thus be ignored.

First, we simulated the FRET efficiency data according to example 1 of Section 5.1. In this irreversible two-state model the lifetime of state 1 is exponentially distributed with an average lifetime of $\lambda^{-1}$ and the conformation-dependent FRET efficiency distributions are a Gaussian distribution for both states. In the simulation, we chose a lifetime of 100 s, so $\lambda = 0.01$ s$^{-1}$. The average FRET efficiency of state 1 and 2 were set to 0.5 and 0.7, respectively, and the standard deviation to 0.07 and 0.06 for state 1 and 2, respectively. This model is plotted in Figure 4A-B. In Figure 5A, the simulated dataset is shown. To extract the parameters from the simulated data, we maximized the likelihood function by using the EM algorithm as described in Section 5.1. The initial parameters were set to $\lambda^{(0)} = 0.025$ s$^{-1}$, $\mu_1^{(0)}=0.35$, $\mu_2^{(0)}=0.9$ and $\sigma_1^{(0)}=\sigma_2^{(0)}=0.12$ and were updated until $Q(\Omega^{(k+1)}|\Omega^{(k)})-Q(\Omega^{(k)}|\Omega^{(k)})$ was smaller than 10$^{-5}$. In Figure 5B-D, the parameters values during each iteration of the EM algorithm are depicted and shows convergence to values close to the true parameters. To compare the model fit with the data, we constructed 3 FRET efficiency histograms over a time interval of 100 s (Figure 5E). The histograms agree well with the model prediction of Eq. \ref{fit2}.
\begin{figure}
\centering
\includegraphics[width=1\textwidth]{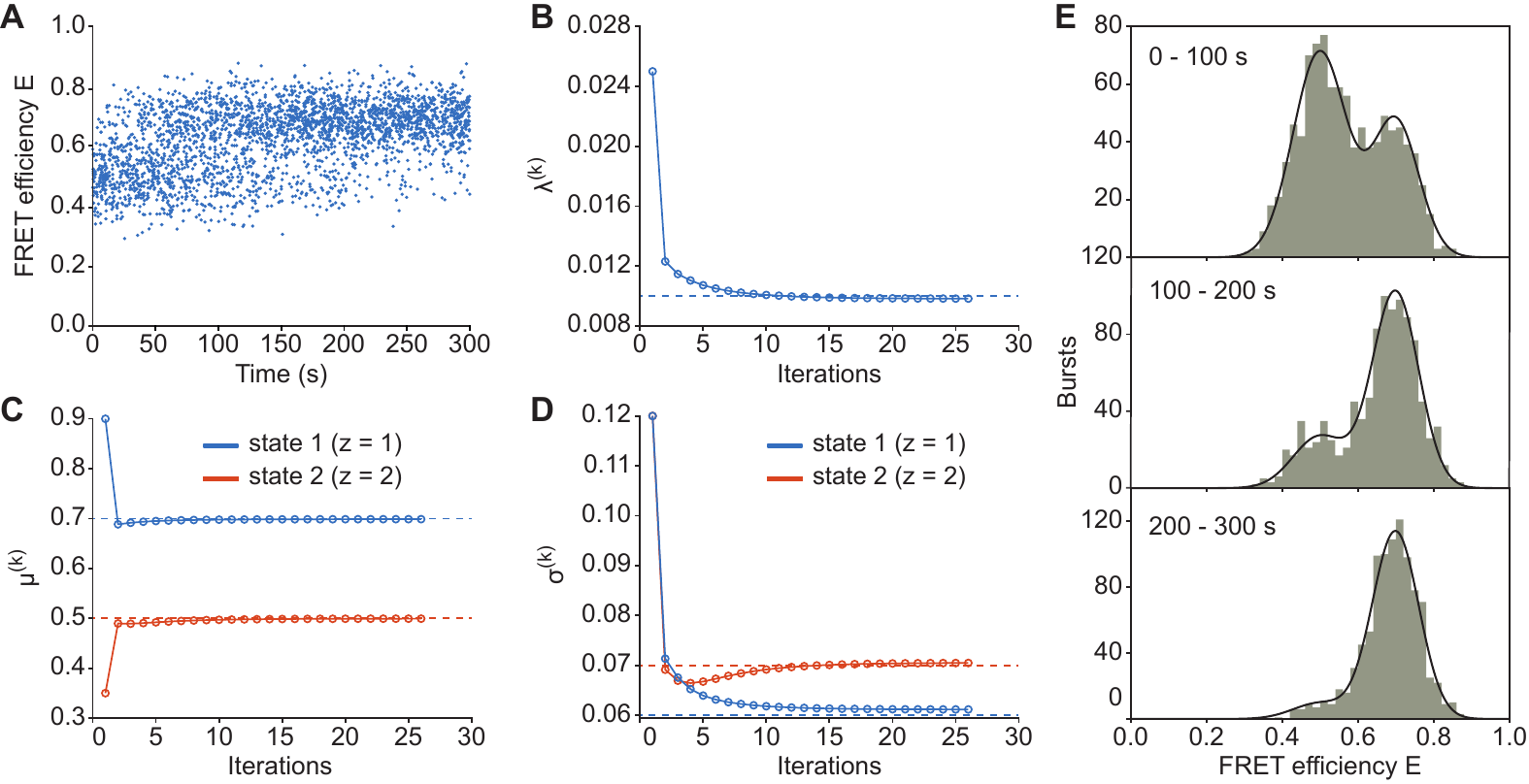}
\caption{(A) Simulated dataset according to the model of Section 5.1, with $\lambda=0.01$ s, $\mu_1=0.5$, $\sigma_1=0.07$, $\mu_2=0.7$ and $\sigma_2=0.06$. Details of the simulation can be found in Section 7. (B-D) The estimated parameters at the $k$th iteration (empty circles) converge to values close to the true parameters (dotted lines). (E) FRET efficiency histograms of the dataset as shown in panel A for the bursts detected within the time interval $\mathcal{T}_1 = [0$ s$,100$ s$]$ (top), $\mathcal{T}_2 = (100$ s$,200$ s$]$ (middle) or $\mathcal{T}_3 = (200$ s$,300$ s$]$ (bottom). The solid line depicts the model prediction as given by Eq. \ref{fit2} evaluated with the estimated parameters.}
\label{fig4}
\end{figure}

To evaluate in more detail the performance of the maximum likelihood method, we simulated 100 datasets, each containing data of 3000 bursts. The average and standard deviation of the estimated parameters are shown in Table 1, which shows that the maximum likelihood estimates indeed converge to the true parameters with a typical error (standard error of the mean) of $<1\%$. 

In Table 2 and 3, the simulation results are shown for alternative models of the conformational dynamics. The data of Table 2, is based on Eq. \ref{model2} with $\lambda = 0.01$ s$^{-1}$ and $\pi = 0.8$ and Table 3 is based on Eq. \ref{model3} with $\tau = 50$ s. The performance of the maximum likelihood method was evaluated by simulating 100 datasets, each consisting of 3000 bursts. The EM algorithm, as described in Section 5.2 and 5.3, was used to estimate the model parameters from the simulated data. Table 2 and 3 show that the maximum likelihood estimates are in perfect agreement to the true parameters. Thus, the simulations indicate that the maximum likelihood method can be used to efficiently obtain the model parameters from the data.

\begin{table}
\renewcommand{\arraystretch}{1.4}
\centering
\caption{Estimated parameters of model Section 5.1}
\smallskip
\begin{tabular}{c|c|c|c}
%\hline
Parameter & Initial parameter & Average  (s.d.) estimated parameters & True parameter  \\
\hline
$\lambda$ & 0.0025 &0.0100  (0.0004) & 0.01 \\
$\mu_1$ & 0.35 &0.4998  (0.0031) & 0.5 \\
$\mu_2$& 0.9&0.6999  (0.0017) & 0.7 \\
$\sigma_1$& 0.12&0.0699  (0.0023) & 0.07 \\
$\sigma_2$& 0.12&0.0600  (0.0012) & 0.06 \\
%\hline
\end{tabular}
\end{table}

\begin{table}[h]
\renewcommand{\arraystretch}{1.4}
\centering
\caption{Estimated parameters of model Section 5.2}
\smallskip
\begin{tabular}{c|c|c|c}
%\hline
Parameter & Initial parameter & Average  (s.d.) estimated parameters & True parameter  \\
\hline
$\lambda$ & 0.025 &0.0100  (0.0004) & 0.01 \\
$\pi$ & 0.4& 0.8032   (0.0321) & 0.8 \\
$\mu_1$ & 0.35& 0.5002   (0.0042) & 0.5 \\
$\mu_2$& 0.9&0.6999   (0.0015) & 0.7 \\
$\sigma_1$& 0.12&0.0697   (0.0031) & 0.07 \\
$\sigma_2$& 0.12&0.0599  (0.0010) & 0.06 \\
%\hline
\end{tabular}
\end{table}

\begin{table}[!h]
\renewcommand{\arraystretch}{1.4}
\centering
\caption{Estimated parameters of model Section 5.3}
\smallskip
\begin{tabular}{c|c|c|c}
%\hline
Parameter & Initial parameter & Average  (s.d.) estimated parameters & True parameter  \\
\hline
$\tau$ & 200 & 50.356  (3.284) & 50 \\
$\mu_1$ & 0.35 & 0.4997   (0.0040) & 0.5 \\
$\mu_2$ & 0.9 & 0.7002   (0.0021) & 0.7 \\
$\sigma_1$ & 0.12 & 0.0695   (0.0025) & 0.07 \\
$\sigma_2$ & 0.12 & 0.0599  (0.0012) & 0.06 \\
%\hline
\end{tabular}
\end{table}

\section{Concluding remarks }
In this paper, we present a framework for the analysis of non-equilibrium solution-based smFRET data. The framework enables obtaining parameter estimates for the conformation-dependent FRET efficiency distribution as well as the distribution that describes the non-equilibrium conformational dynamics. Parameter estimates were found by maximizing the appropriate likelihood function. By making use of the EM algorithm, the maximization task is significantly simplified. We provided illustrative examples of models for the conformational dynamics, which represent various simple non-equilibrium processes. More complex models could be implemented in a similar way.

Information about the conformational dynamics of single-molecules has predominately been obtained through the analysis of immobilized molecules. In these surface-based smFRET measurement, individual molecules are attached to the surface of a glass slide and followed over time by using confocal scanning microscopy \cite{nsmb} or total internal reflection microscopy \cite{erkens}. This data could give direct insight into the order of events and the kinetics associated with each step. However, one disadvantage of surface-based measurements is that the molecules need to be specifically immobilized without effecting their function, which can sometimes be challenging, especially for large macromolecular complexes. Immobilization might also be inappropriate if processes like protein folding and chaperone interactions are studied. Besides the technological developments in microfluidic mixing to obtain dynamic information from freely-diffusing molecules with smFRET \cite{pmid23845960, pmid19485532}, information about the dynamics can also be obtained with conventional solution-based smFRET measurements as considered in this work \cite{NCP,abce1,pmid31356875,dna_motor,dna_motor2}. Analysis methods for solution-based smFRET measurements have been developed to extract information about fast conformational dynamics, which would occur on timescales similar to the residence time of the molecule inside the excitation volume (around 1 ms) \cite{bva, pda, pmid17929964}. The analysis framework provided in this paper would be appropriate to study slower processes, which would occur on or above the sub-minute timescale. This limit might be significantly lowered due to the recent development of high-throughput smFRET systems \cite{pmid31356875}. 

We note that alternative analysis methods can also be used. For instance, a histogram for the FRET efficiencies could be constructed for different non-overlapping time intervals (see also Section 6). The histograms could then be fitted with a time-independent GMM (Section 5.4), similarly to what would be done in an equilibrium situation. One could equate the estimates for the mixture weights to the probability of acquiring the corresponding conformational states at a timepoint that is equal to the center of the time interval. However, this approximation would only be valid if the distribution to acquire the different conformational states is approximately constant over the time interval. If this approximation can be made, a course-grained picture of the conformational dynamics can be obtained, which can be further analyzed. However, because a sufficient number of observations are needed to properly estimate the mixture weights, a minimum duration for each time interval would be around 3 min for simple systems. Therefore, this time interval method would only be appropriate to analyse very slow processes that occur on time scales much slower than 3 min (say 20 min or above).

This work focuses on obtaining parameter estimates from the data by maximizing the likelihood function. However, further work is needed to establish the errors on those estimates. In principle, multiple sources of error or uncertainty should be considered. First, if the model for the data is incorrect, the wrong likelihood function is maximized. This is the subject of model selection. It would be good practice to test different (nested) models for the data and select the most appropriate model by, for example, the likelihood ratio test. Alternatively, the BIC or the closely related AIC could be used for model selection. Moreover, the discrepancy between the model prediction and the data could be examined as described in Section 6. Secondly, errors can arise because of the limited amount of data. A way to deal with this type of uncertainty is by bootstrapping or by evaluating the Fisher Information. Thirdly, the EM algorithm might converge to a local maximum of the likelihood function instead of the global maximum. Varying the starting conditions of the algorithm might help in finding the global maximum. 

Maximum likelihood estimators have the desired invariance property, i.e., if $\hat{\theta}$ is the maximum likelihood estimator for $\theta$, then $g(\hat{\theta})$ is the maximum likelihood estimator for $g(\theta)$. This property can be used to estimate other physical properties from the data. For instance, once the maximum likelihood estimates for the average FRET efficiencies are obtained, i.e. $\mu_z$ in case of a Gaussian distribution, then the maximum likelihood estimates for the donor-acceptor distance $r$ could obtained by solving $\mu_z=  R_0^6 / (R_0^6+r^6 ) $. This could give important insight in to the molecular structure of the conformational states, especially if multiple measurements are combined with different donor and acceptor labeling positions. 

\section{Acknowledgements}
This work was done in the Poolman lab at the University of Groningen. I would like to thank Monique Wiertsema and Bert Poolman for critically reading of the manuscript. I also would like to thank Monique Wiertsema for the help with the simulations.

\bibliography{ms}
\bibliographystyle{ieeetr}

\end{document}